\documentclass[conference]{IEEEtran}
\IEEEoverridecommandlockouts
\usepackage{cite}
\usepackage{amsmath,amssymb,amsfonts}
\usepackage{algorithmic}
\usepackage{graphicx}
\usepackage{textcomp}
\usepackage{xcolor}
\usepackage{caption}
\usepackage{subcaption}
\usepackage[algo2e,ruled,linesnumbered]{algorithm2e}
\usepackage{balance}
\usepackage{booktabs}

\usepackage[english]{babel} 
\usepackage{amsthm}
\theoremstyle{definition}
\newtheorem{definition}{Definition}[section]

\usepackage{enumitem} 

\newcommand{\eg}{\textit{e.g.,}}

\newcommand{\etal}{\textit{et. al. }}

\begin{document}

\title{RT-DBSCAN: Accelerating DBSCAN using Ray Tracing Hardware\\
\thanks{We thank the anonymous IPDPS reviewers for their valuable feedback. We thank Dr. Eleazar Leal for providing the G-DBSCAN code and Dr. Andrey Prokopenko for answering our questions about FDBSCAN. We are grateful to Kirshanthan Sundararajah for his insightful comments that helped improve the paper. This work was funded by NSF grants CCF-1908504, CCF-1919197 and CCF-2216978.}
}
\author{\IEEEauthorblockN{Vani Nagarajan}
\IEEEauthorblockA{
\textit{Purdue University, USA} \\
nagara16@purdue.edu}
\and
\IEEEauthorblockN{Milind Kulkarni}
\IEEEauthorblockA{
\textit{Purdue University, USA} \\
milind@purdue.edu}
}
\thispagestyle{plain}
\pagestyle{plain}
\maketitle
\begin{abstract}
General Purpose computing on Graphical Processing Units (GPGPU) has resulted in unprecedented levels of speedup over its CPU counterparts, allowing programmers to harness the computational power of GPU shader cores to accelerate other computing applications. But this style of acceleration is best suited for {\em regular} computations (e.g., linear algebra). Recent GPUs feature new {\em Ray Tracing (RT)} cores that instead speed up the irregular process of ray tracing using Bounding Volume Hierarchies. While these cores seem limited in functionality, they can be used to accelerate n-body problems by leveraging RT cores to accelerate the required distance computations. In this work, we propose RT-DBSCAN, the first RT-accelerated DBSCAN implementation. We use RT cores to accelerate Density-Based Clustering of Applications with Noise (DBSCAN) by translating fixed-radius nearest neighbor queries to ray tracing queries. We show that leveraging the RT hardware results in speedups between 1.3x to 4x over current state-of-the-art, GPU-based DBSCAN implementations.
\end{abstract}


\section{Introduction}
Graphics Processing Units (GPUs) were created to service graphics applications and accelerate parts of the rasterization pipeline. The acceleration was due to optimized floating point arithmetic using a large number of arithmetic cores. Researchers wanted to harness this massive computational capability to accelerate non-rendering applications. However, this was not straightforward as it required re-formulating problems as 3D rendering problems. Over time, the emergence of platforms such as CUDA \cite{cuda} and OpenCL\cite{open-cl} provided a programming model for general-purpose computations on GPUs. The General-purpose GPU (GPGPU) programming model lets users offload parallelizable and compute-intensive components (\eg, training a neural network) to the GPU by leveraging the programmable shader cores. Unfortunately, GPGPU acceleration using shader cores remains largely the province of {\em regular} applications that rely on dense loops over dense structures, such as matrix operations, convolutions, etc.

\paragraph{GPU acceleration of ray tracing}
Interestingly, while GPU shader cores are well-suited for one style of graphics rendering---rasterizing---they are not at all well suited to a different, more accurate, rendering algorithm, {\em ray tracing}.
Ray tracing (or, more accurately, {\em ray casting}) flips the approach of rasterizing. Rather than considering each object and how it affects one or more pixels on the screen (e.g., whether the object is visible or occluded by other objects), ray tracing considers each {\em pixel} in the scene and determines what color it should be based on the objects and lights it interacts with \cite{ray_trace_whitted}.
In ray tracing, a ray is cast from the source, which is typically a pinhole camera, for every pixel in the image plane. 
These primary rays pass through the image plane and their interactions with objects in the scene determine the color of the pixel. The ray could intersect an object in the scene, creating new reflected, refracted and/or shadow secondary rays. This hierarchy of rays is represented as a ray tree, with the primary ray as the root and the spawned secondary rays as internal nodes (secondary ray intersections can spawn tertiary rays) or leaf nodes.

Ray-object intersection tests are the biggest bottleneck in the ray tracing pipeline due to their computational intensity. As {\em each} ray has to be tested for intersection against {\em every} object in the scene, performance suffers greatly.
However, it is not {\em always} necessary to test for intersection against every object. We can represent the objects in the scene using a Bounding Volume Hierarchy (BVH), a type of {\em spatial acceleration tree} \cite{waldbvh}. A BVH, like other spatial trees, captures the relationship of objects to each other in space by recursively subdividing the space into smaller cells until the leaf cells contain bounding volumes that contain single objects. Ray-object intersection can thus be performed hierarchically: if a ray does not intersect a bounding volume, then it cannot intersect any of the subordinate bounding volumes in the tree, eliminating large numbers of intersection tests. (Section~\ref{sec:bvh} describes this process in more detail.)

Unfortunately, BVH-based ray-tracing is {\em highly} irregular: each ray performs a tree traversal whose extent is highly input-dependent. While prior work has shown that tree traversals can be performed reasonably well on GPGPUs~\cite{goldfarb13sc}, shader cores are simply not the best-suited accelerator for BVH traversals. Hence, recent GPUs from NVIDIA and AMD have added {\em ray tracing (RT) cores}. These cores provide specialized hardware for building and traversing bounding volume hierarchies, significantly accelerating the process of ray tracing \cite{whitepaper}.

\paragraph{Re-purposing RT cores}
In the same way that early GPU programmers looked to re-purpose shader cores to perform non-rasterization tasks, a natural question to ask is whether RT cores can be leveraged to accelerate non-raytracing algorithms. Recent work has suggested the answer might be ``yes.'' Wald~\etal~\cite{wald19} accelerate the task of locating which tetrahedron of a solid a query point lies in by treating the query point as the source of a ray and seeing if it intersects a tetrahedron. While this is perhaps an obvious adaptation of ray-tracing, as it is effectively ray tracing itself, later work has pushed the boundaries further. Zellman~\etal~\cite{forcegraph} and Evangelou~\etal~\cite{Evangelou2021RadiusSearch} showed how to reduce the problem of finding the set of points in a fixed-radius neighborhood of a query point to ray tracing to build force-directed graphs and to do photon mapping, respectively.\footnote{This reduction is described in more detail in Section~\ref{sec:design}.} 

These papers use this reduction to write custom ray-tracing kernels that solve these problems. The algorithms essentially use one-shot invocations of ray-tracing hardware to perform the necessary distance computations and solve the problems. However, some distance-based algorithms require integrating repeated distance queries into larger programs, and hence require more careful use of the reduction. 

In particular, DBSCAN \cite{Ester96adensity-based} is a clustering algorithm that groups nearby points in space into clusters based on the distance points within the cluster are from other points in the cluster. Solving this problem requires repeatedly updating cluster definitions by repeatedly identifying nearby points, a more complex task than the ``single shot'' distance computations of prior RT-acceleration work. In this paper, we investigate whether RT cores can be used to accelerate this algorithm. The contributions of our paper are as follows:
\begin{itemize}
    \item This paper introduces RT-DBSCAN, the first RT-accelerated clustering algorithm. As DBSCAN uses distance-based queries to identify neighboring points, we are able to leverage the reduction of Zellman~\etal~\cite{forcegraph} and Evangelou~\etal\cite{Evangelou2021RadiusSearch} to accelerate neighbor searches, which are a major computational bottleneck.
    \item We create a primitive {\tt RT-FindNeighbor} (Details in Section~\ref{alg:rt-neigh}), allowing us to easily negotiate with the ray tracing hardware and its associated programming model (the Optix Wrapper Library (OWL)~\cite{owl}).
    \item  We use {\tt RT-FindNeighbor} to implement a {\tt UNION-FIND}-based DBSCAN algorithm (Details in Section~\ref{alg:parallel_dbscan}) that minimizes memory consumption. We proceed to show that RT-DBSCAN outperforms current state-of-the-art DBSCAN algorithms that have been optimized to run on GPUs.
\end{itemize}


\section{Background}
\subsection{Spatial Trees} \label{sec:bvh}
n-body problems encompass all problems where every data point needs to interact with all other {\em n} data points to calculate some result. Examples of n-body problems include (1) force-directed graph drawing algorithms such as Spring Embedders\cite{Eades1984AHF}, where repulsive forces between all pairs of vertices are used to find stable graph layouts, (2) cosmology simulations, where the gravitational force exerted by stars in a galaxy is modeled, and (3) k-nearest neighbors \cite{Altman1992AnIT}, where the distance between all points in the dataset is calculated to identify the k-nearest neighbors, to name a few. The naive approach to solving n-body problems is to loop over the {\em n} data points and compute the effects of interaction with the other $n-1$ data points in a doubly nested loop, leading to an algorithm with $O(n^2)$ time complexity.

Barnes-Hut\cite{barnes86a} improved the complexity by introducing a tree representation of input data points called Spatial Index Tree. They built the tree by recursively grouping nearby points using spatial subdivision. The individual points formed the leaves of the tree and internal nodes represented groupings that estimated the effect of the points contained in them. With this representation, instead of computing the effect of each point on {\em all} other points, we compute the effect of a {\em group} of points on other {\em individual} points. Using this spatial tree optimization, the time complexity of n-body problems reduced from $O(n^2)$ to {\em O(n log n)}.

\subsubsection{Bounding Volume Hierarchy}
The grouping of points can be done by either spatial (R-Trees, Oct-trees) or object-based subdivision of the dataset (Bounding Volume Hierarchies). Ray tracing applications rely on Bounding Volume Hierarchies (BVH) to reduce the number of ray-object intersection tests performed to determine the coloration of a pixel. Analysis of ray tracing execution times shows that about 70\% of the total time is spent testing for intersections for simple scenes, and upto 95\% for complex scenes \cite{fuji}. The authors also state that the main reason for these percentages is the {\em number} of intersection tests that need to be performed. 

\subsubsection{BVH Build}
\begin{figure}
     \centering
     \begin{subfigure}[t]{0.45\textwidth}
         \centering
         \includegraphics[scale=0.5]{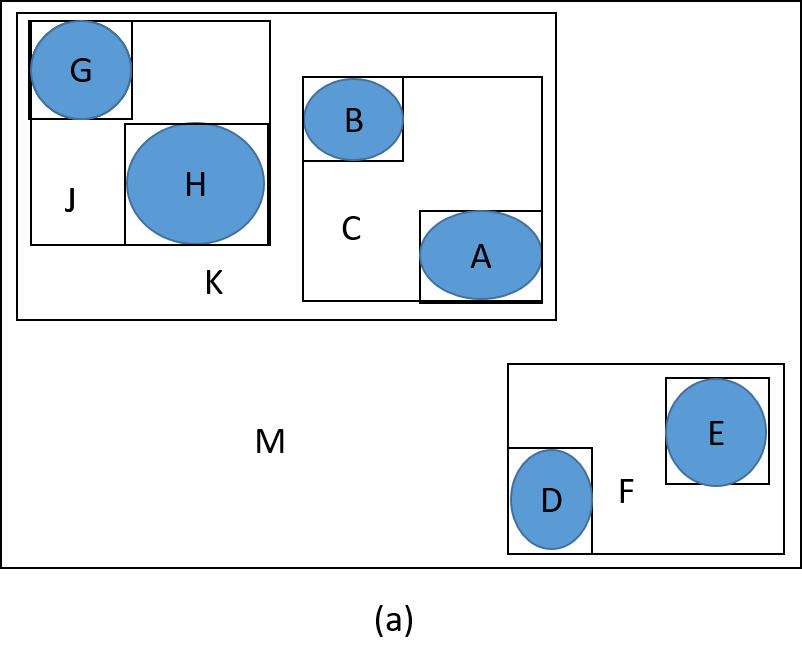}
         \vspace{-0.5em}
         \caption{2D rectangular bounding boxes for the objects and intermediate bounding volumes}
         \label{fig:bvh_img_a}
     \end{subfigure}    
     \hfill
     \begin{subfigure}[t]{0.45\textwidth}
         \centering
         \includegraphics[scale=0.5]{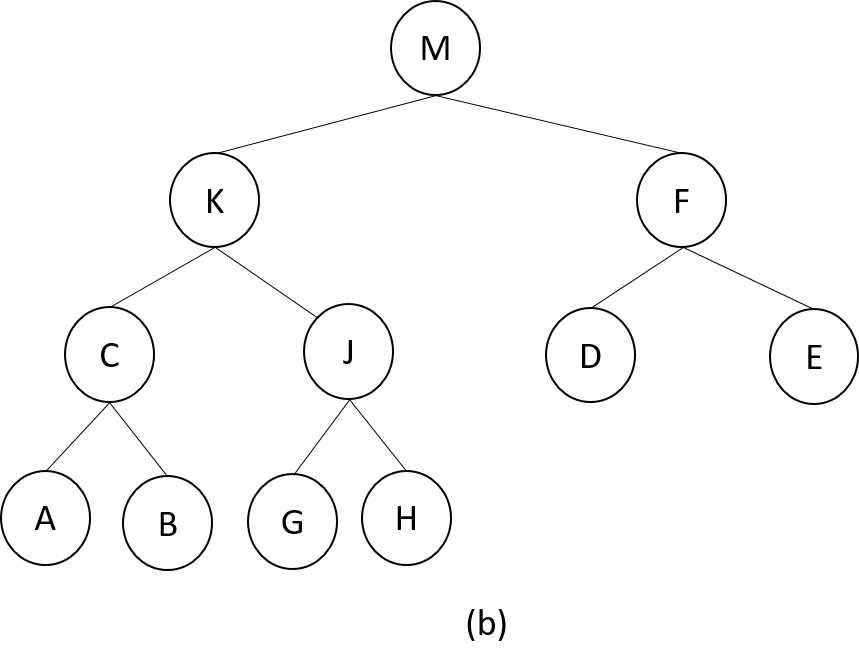}
         \vspace{-0.5em}
         \caption{Bounding Volume Hierarchy built from the bounding boxes in (a)}
         \label{fig:bvh_img_b}
     \end{subfigure}
     \caption{Bounding Volume Hierarchy Construction}
     \label{fig:bvh_img}
     \vspace{-0.5em}
\end{figure}

In the initial stage of the ray tracing algorithm, all objects in the scene are enclosed in bounding volumes. It is possible for the bounding volumes to overlap if the objects are sufficiently close. A {\em bounding volume} is a closed volume that completely encompasses one or more objects. We use Axis-Aligned Bounding Boxes (AABBs) as the bounding volumes. The BVH is built bottom-up, starting with the leaves. Bounding volumes containing individual objects in the scene are the leaf nodes of the tree. Fig~\ref{fig:bvh_img_a} shows objects {\em A, B, G, H, D} and {\em E} enclosed in rectangular (shown in 2D) bounding boxes and Fig~\ref{fig:bvh_img_b} shows the corresponding bounding boxes as leaves of the tree. The bounding volumes containing individual objects are then recursively grouped together to create larger bounding volumes, forming internal tree nodes {\em C, J, F} and {\em K}. This process continues until a single bounding box, {\em M}, encloses every intermediate and individual bounding volume. In Fig~\ref{fig:bvh_img_a}, intermediate bounding volume {\em C} encloses bounding volumes {\em A} and {\em B}, and {\em J} encloses {\em G} and {\em H}. {\em C} is combined with bounding volume {\em J} to create {\em K}, which is further combined with {\em F} to create {\em M}, which encloses the entire scene. The corresponding hierarchical relationship is captured in Fig~\ref{fig:bvh_img_b}, with {\em C} as the parent of {\em A} and {\em B}, {\em J} as the parent of {\em G} and {\em H}, {\em K} as the parent of {\em C} and {\em J}, and {\em M} being the parent of {\em K} and {\em F}. 

\subsection{Ray Tracing Cores}\label{sec:rtx}
The addition of Ray Tracing (RT) cores to GPUs has enabled hardware-accelerated real-time ray tracing in gaming applications. These accelerators co-exist with the traditional Streaming Multi-processors (SMs), and the Optix API (Section~\ref{sec:optix-api}) allows us to write code that can leverage both RT and shader cores of the GPU. The RT cores\cite{whitepaper} accelerate Bounding Volume Hierarchy traversal and ray-triangle intersection tests, which are crucial and expensive elements of the ray tracing pipeline. 

\subsubsection{BVH Traversal}\label{sec:bvh-traversal}
Given objects in the scene, the RT cores intelligently\footnote{Details of the actual hardware internals are not publicly available} build a Bounding Volume Hierarchy, similar to the description in Section~\ref{sec:bvh}. The reduction in the number of intersection tests performed comes from pruning large parts of the search space by performing intersection tests on {\em bounding volumes} rather than individual objects. During BVH traversal, if a ray does not intersect a bounding volume, it {\em cannot} intersect any of the volumes contained in it and traversal does not continue down that subtree. In Fig~\ref{fig:bvh_img}, if a ray does not intersect bounding volume {\em K}, we need not test for intersection against {\em C}, {\em J}, {\em A}, {\em B}, {\em G} or {\em H}.

\subsubsection{Optix Programming Model}\label{sec:optix-api}
The Optix API \cite{prog-guide} allows users to write custom shader programs in CUDA (processed as a single CUDA kernel), in addition to offloading BVH build and traversal to RT cores. Both the shader and RT cores in GPUs use the same device memory and work can be done in parallel on the two cores. If the GPU does not have an RT core, Optix programs can still be run, with BVH build and traversal being performed in software. 

Optix\footnote{All Optix kernels are also present in OWL\cite{owl}} allows users to set up their scene by providing support for triangles, spheres and other user-defined geometries. Once the scene is set up, the user can define a bounding volume program to enclose objects in the scene. For geometries such as spheres, axis-aligned bounding boxes are typically used. The bounding volumes are recursively combined by the RT cores to build the BVH, as explained in Section~\ref{sec:bvh}. With the BVH constructed, we can now create rays and trace their interactions with objects in the scene by traversing the BVH and performing intersection tests.

\begin{figure}[t]
    \centering
    \includegraphics[width=\linewidth]{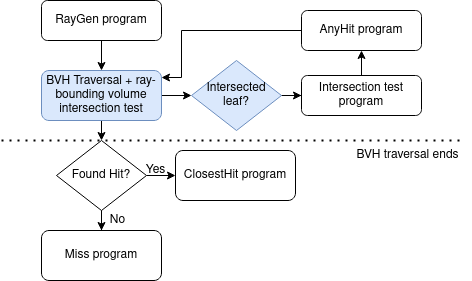}
    \caption{Optix pipeline. The components shaded in blue are performed in hardware and are not programmable. The unshaded components can be defined by the user. In the case where objects are triangles, the Intersection Test is also performed in hardware}
    \label{fig:optix_api}
    \vspace{-1.5em}
\end{figure}

As ray tracing is an {\em embarrassingly parallel} problem where the color of each pixel can be computed independently, Optix allows multiple rays to be launched in parallel as separate CUDA threads. 
The components of the Optix pipeline are shown in Fig~\ref{fig:optix_api} and each ray executes the various stages in parallel. The {\tt RayGen} program generates parallel rays with the given origin ($\Vec{o}$) and direction ($\Vec{d}$). The user also needs to specify the ray interval $[t_{min}, t_{max}]$, which determines the extent of the ray:
\[\Vec{r} = \Vec{o} + t\Vec{d}, t \in [t_{min}, t_{max}]\]

The generated rays both traverse the BVH and test for intersection with bounding volumes in hardware (shown in blue in Fig~\ref{fig:optix_api}). When the ray reaches a {\em leaf} bounding volume enclosing an object, the ray-{\em object} intersection test is performed in software or hardware, depending on the object. If the object is a triangle, the test is performed in hardware. Otherwise, the user specifies the {\tt Intersection} program to be used for ray-object intersection testing. The user can optionally specify an {\tt AnyHit} program to record information about all the intersected objects and to determine whether to continue or terminate BVH traversal. Once the BVH traversal has completed, the user can optionally call the {\tt ClosestHit} program to identify the object closest to the ray along its path or the {\tt Miss} program to handle cases with no ray-object intersections.

\subsection{Density-Based Clustering of Applications with Noise} \label{sec:dbscan}
Cluster analysis is an unsupervised learning technique used to identify patterns in a dataset. Clustering techniques are widely used to provide targeted advertising to customers with similar purchase histories, identify faulty machines, detect anomalous financial transactions, and so on. {\em k}-means \cite{kmeans}, a popular clustering technique due to its simplicity and scalability, picks {\em k} centroids and forms clusters based on whether other points in the dataset are within a permissible distance to the centroids. However, it requires that the user specify the number of clusters ({\em k}) to be formed and performs poorly when the dataset is noisy. Density-Based Clustering of Applications with Noise (DBSCAN) addresses the disadvantages of {\em k}-means, as it does not require the user to specify the number of clusters to be formed\cite{Ester96adensity-based}. It has the added advantages of being able to form clusters of varying shapes and being unperturbed by noisy datasets. 

The DBSCAN algorithm takes two parameters as inputs: $\varepsilon$ and {\em minPts}, where $\varepsilon$ is the maximum permissible distance between any two points in a cluster and {\em minPts} is the minimum number of points within the $\varepsilon$-neighborhood required to form a cluster. A point is said to be a {\em Core Point} if it has {\em minPts} neighbors within $\varepsilon$ distance. A {\em Border Point} is not a core point but is {\em reachable} from a core point and is a part of a cluster. Reachability comes in two forms: (1) {\em directly reachable}, where point y is within a distance $\varepsilon$ from core point x, and (2) {\em reachable}, where y is connected to core point x through one or more core points. A {\em Noise Point} is neither a core nor a border point.  
\begin{algorithm2e}
\caption{Original DBSCAN}\label{alg:dbscan}
\For{UNASSIGNED point p} {$Neighbors \gets FindNeighbors(p)$ \\
\eIf{$Neighbors.length < minPts$}
{
    $p \gets NOISE$ \\
}{
    $p \gets CLUSTER\_ID$ \\
    $NeighborSet \gets Neighbors - \{p\}$ \\
    \For{$neighbor \in NeighborSet$} 
    {
        \If{neighbor == UNASSIGNED $\|$ \\ neighbor == NOISE} 
        {
            $neighbor \gets CLUSTER\_ID$ 
            $NewNeighbors \gets FindNeighbors(neighbor)$ \\
            \If{$NewNeighbors.length \ge minPts$} 
                {$NeighborSet \gets NeighborSet \bigcup NewNeighbors$}
        }
    }
}
}
\end{algorithm2e}

Algorithm~\ref{alg:dbscan} shows the original DBSCAN algorithm. Initially, all points are considered UNASSIGNED as they have not been assigned to a cluster yet. The {\em FindNeighbors(p)} function in Line 2 identifies all points within $\varepsilon$ distance of {\em p}. In lines 3-6, we classify the point as a Core point or a Noise point based on whether the $\varepsilon$ neighborhood of {\em p} has {\em minPts} points. If the point is a Core point, we examine each neighbor and assign it the same Cluster\_ID as the core point, as shown in Lines 9-11. If the neighbor has already been assigned a Cluster\_ID, we ignore the point and move on to the next neighbor. In lines 13-16, we call the {\em FindNeighbors} function on the {\em neighbors} of point {\em p}. If the neighbor is a Core point, we add it to the set of neighbors and repeat the process until all points have either been assigned to a cluster or classified as noise.

\section{Design} \label{sec:design}
\subsection{Neighbor Search}
In Section~\ref{sec:dbscan}, we introduced an algorithm that performed density-based clustering. Algorithm~\ref{alg:dbscan} includes two references to a {\em FindNeighbors()} function that identifies all points within $\varepsilon$ distance of a point. We generalize the query as follows:

\begin{definition} $\mathit{findNeighborhood}(p, S, \varepsilon)$: 
Given a dataset $S$, point {\em p} and distance $\varepsilon$, find all points $\{x \in S ~|~ \mathit{distance}(p,x) \le \varepsilon\}$
\end{definition}

The $\mathit{distance(x,y)}$ function calculates the Euclidean distance between points $x$ and $y$. The answer to this question is used to establish the $\varepsilon$-neighborhood to detect core points in DBSCAN. In Section~\ref{sec:rtx}, we discussed how RT cores are used to determine the color of a pixel by answering the ray-object intersection query:

\begin{definition} $\mathit{intersect}(\vec{r}, S)$
Given a set of objects $S$ in a scene and ray, $\vec{r}$, find all 
objects $\{o \in S ~|~ \vec{r}$ $\mbox{intersects}$ $o\}$
\end{definition} 

The key question, then, is to find a way to implement $\mathit{findNeighborhood}$ in terms of $\mathit{intersect}$. If we can do this, then algorithms such as DBSCAN can be readily written in terms of $\mathit{intersect}$ and can use the RT cores to accelerate their execution. We describe this process next, adapting a reduction proposed by prior work~\cite{Evangelou2021RadiusSearch,forcegraph}.

\subsection{Input Transformation}\label{sec:input_trans}
\begin{figure}[t]
    \centering
    \includegraphics[width=0.7\linewidth]{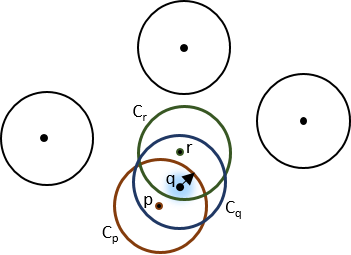}
    \caption{We expand spheres of radius ($\varepsilon$) around all points. We launch an infinitesimally small ray originating at query point {\em q} and see that it intersects $C_p$, $C_q$ and $C_r$, since the origin is contained within all 3 spheres. Hence, {\em p, q} and {\em r} are {\em q}'s neighbors}
    \label{fig:sphere_expand}
    \vspace{-1.5em}
\end{figure}

 The key insight to the transformation is that points within a distance $\varepsilon$ of a query point {\em p} (within {\em p}'s $\varepsilon$-neighborhood) are the {\em same} points that would be contained inside a sphere with origin {\em p} and radius $\varepsilon$. This intuitively makes sense for 3D datasets, since expanding a sphere over the query point would span across the three dimensions and accumulate {\em all} points within a particular query radius. However, Zellmann~\etal \cite{forcegraph} propose an alternate approach where, instead of expanding a sphere only around the query point {\em p}, they expand spheres of radius $\varepsilon$ around {\em all} points in the dataset. Fig~\ref{fig:sphere_expand} shows how spheres $C_p$, $C_q$ and $C_r$ are expanded over points {\em p}, {\em q} and {\em r}. We notice that the spheres overlap if their centers are in each other's $\varepsilon$ neighborhood. 

\subsection{Putting it all together: RT-accelerated $\mathit{findNeighborhood}$}\label{sec:rt-findNeigh}
Now that we have a way of representing our points as spheres in a scene, we need to formulate our ray tracing query such that we can identify all neighbors of a point. 
Algorithm~\ref{alg:rt-neigh} presents a high-level overview of the process. 

The algorithm takes a query point {\em q}, query radius $\varepsilon$ and the dataset {\em D} as inputs. In lines 1-3, for each point $p_i$, we add a sphere primitive with origin $p_i$ and radius $\varepsilon$ to the scene. This is the input transformation process described in Section~\ref{sec:input_trans} and shown in Fig~\ref{fig:sphere_expand}. 

Using the user-specified axis-aligned bounding box program, a Bounding Volume Hierarchy is constructed in hardware to create the scene. In Line 4, we launch an infinitesimally small ray $\vec{r}$ with origin $\vec{q}$, direction $\vec{d}$ and $[t_{min}, t_{max}]$ as $[0, 1e^{-16}]$ to find all the spheres that are intersected by the ray. For example, from Fig~\ref{fig:sphere_expand}, we see that such a ray launched from origin  $\vec{q}$ intersects $C_p, C_q$ and $C_r$. Recall that these are solid 3D spheres and a ray of infinitesimal length is sufficient to find intersections with overlapping spheres. The overlap of spheres in Fig~\ref{fig:sphere_expand} indicates that the centers of the spheres $C_p$ and $C_r$ are within $\varepsilon$ distance of $\vec{q}$. 

In Lines 5-9, we record all the spheres intersected by the ray and pass it through a filter to remove self-intersections ($C_q$ in this case). When the ray traverses the BVH in hardware, it returns all the intersected {\em bounding volumes} in the BVH tree. 
As the hardware tests for intersection with {\em bounding volumes} and not {\em objects}, it is possible that the intersection test results are incorrect. Though bounding volumes completely enclose objects, they are not an {\em exact} fit. It is possible for the ray to intersect the bounding volume but completely miss the object contained in it. In Line 6, we perform an additional check to confirm that we intersect the {\em object}. Since it is also possible for bounding volumes to overlap if the dataset is dense, this filter removes any erroneous bounding volume intersections. The filtered list of intersected spheres ({\em NeighborList}) contains the nearest neighbors of point {\em q}.

\begin{algorithm2e}
\caption{RT-FindNeighborhood}\label{alg:rt-neigh}
\SetKwInOut{Input}{Input}
\SetKwInOut{Output}{Output}
\Input{Query point q, Query radius $\varepsilon$, Dataset D}
\Output{NeighborList}
\For{$p \in D$} {$S \gets S \bigcup createSphere(p, \varepsilon)$\\  
} 
traceRay($\vec{q}, \vec{d}, [t_{min}, t_{max}]$)\\
\If{Intersect(q, $s \in S$)}{
\If{$(dist(q, s) \le \varepsilon) \wedge (q \ne s)$}{$NeighborList \gets NeighborList \bigcup s$}}
\end{algorithm2e}
\vspace{-1.0em}
\begin{algorithm2e}
\caption{RT-DBSCAN}\label{alg:parallel_dbscan}
\For{point p} {$neighborCount \gets RT$-$FindNeighbors(p).length$ \\
    \If{neighborCount $\ge$ {\em minPts}}
    {
        $p \gets CORE\_POINT$ \\
    }
}
\For{point p} {
   \For{n: RT-FindNeighbors(p)} {
        \eIf{n == CORE\_POINT}{{\tt Union}(p,n)}
        {\If{n == UNCLASSIFIED}{critical section:\\ {\tt UNION}(p,n)}}
    
    }
}
\end{algorithm2e}

\subsection{RT-DBSCAN} \label{sec:rt-dbscan}
We base our {\tt RT-DBSCAN} algorithm on the parallel Union-Find FDBSCAN algorithm proposed by Prokopenko~\etal \cite{DBLP:journals/corr/abs-2103-05162}. Algorithm~\ref{alg:parallel_dbscan} has two stages: (1) identifying Core points, and (2) updating cluster information using {\tt union-find}. For the first stage, we use Algorithm~\ref{alg:rt-neigh} to identify each point's {\em neighbors}. For each point in the dataset, we launch a ray tracing query to check if the ray intersects more than {\em minPts} spheres. If the ray {\em does} intersect more than {\em minPts} spheres, the query point is marked as a Core point as shown in Lines 3-5.

In the second stage, we begin to form the clusters. As we did not save information about the neighbors of each point, we re-compute the Euclidean distance between points in the dataset using Algorithm~\ref{alg:rt-neigh}. Though this computation is redundant and may seem inefficient, the hardware-accelerated BVH traversal prevents performance degradation. In fact, Prokopenko~\etal show that performance does not suffer even without hardware acceleration\cite{DBLP:journals/corr/abs-2103-05162}. Additionally, this approach scales well to larger datasets as we do not run out of memory.

In Lines 7-9, we check whether the point and its neighbor are core points. If both are core points, they can be merged to form a larger cluster using the {\tt UNION} operation. If the neighbor is not a core point, it must be a border point and we merge the border point into the core point's cluster. It is necessary to perform Line 14 atomically as border points can belong to more than one cluster. If not, the border point could be incorrectly assigned to two clusters, causing the erroneous merging of two different clusters. We use a DisjointSet \cite{Hopcroft1973SetMA} structure to store the rank and parent of the point. At the end of the second stage, all points that share the same parent are a part of the same cluster and all other points are noise.

\section{Implementation} 
We implemented RT-DBSCAN using the Optix Wrapper Library (OWL), which is built on top of Optix 7. The tests were run on an NVIDIA GeForce RTX 2060 GPU (with RT cores) with 6 GB device memory, CUDA version 10.1 and Optix 7.1.

OWL has separate programs for different components of the ray tracing pipeline: bounding box construction, intersection test, closest hit and any hit. We implemented the fixed-radius nearest neighbors search and the DBSCAN clustering phases within the Intersection program, saving the cost of calling the {\tt AnyHit} or {\tt ClosestHit} program.

 As Optix only accepts 3D inputs, we set the z-dimension to 0 for 2D datasets and set the z-dimension of the ray direction as 1. We also explicitly disabled the {\tt AnyHit} and {\tt ClosestHit} programs to avoid overhead costs. 

\section{Evaluation}
\subsection{Datasets}
We use four real-world datasets to evaluate RT-DBSCAN. As RT cores can only handle datasets with at most 3 dimensions, we chose these 2D and 3D datasets that have been widely used to evaluate DBSCAN performance(\cite{cuda-dclust,dbscan-compare,DBLP:journals/corr/abs-2103-05162}). 
\begin{description}
\item[3DRoad]
The 3DRoad dataset was constructed using the road network information of North Jutland, Denmark\cite{3droad}. The dataset consists of {\em 435K} points and we use it as a 2D dataset, considering only the latitude and longitude coordinates.
\item[NGSIM]
The Next Generation Simulation (NGSIM) Vehicle Trajectories dataset provides precise vehicle locations along three US highways\cite{ngsim}. The dataset has more than {\em 11M} points and we use the local coordinates to construct a 2D dataset.
\item[Porto]
The Taxi Service Trajectory-Prediction Challenge dataset collected trajectory data of 442 taxis in the city of Porto, Portugal\cite{porto}. The dataset has just over {\em 1M} points and we use the 2D GPS coordinates to identify clusters.
\item[3DIono]
The 3D Ionosphere dataset describes the behavior of weather in the ionosphere\cite{3diono}. The dataset has just over {\em 1M} points and we construct the 3D dataset using latitude, longitude and total electron count parameters. 
\end{description}

\subsection{Performance Evaluation}
We compare RT-DBSCAN against three GPU-based DBSCAN implementations (though none of these use RT cores). 

\begin{description}
\item[FDBSCAN]
FDBSCAN uses a parallel DisjointSet algorithm to compute clusters\cite{DBLP:journals/corr/abs-2103-05162}. It has minimal memory footprint and uses Bounding Volume Hierarchies to minimize the number of distance computations.

\item[G-DBSCAN]
G-DBSCAN stores $\varepsilon$-neighborhood information for all points in a graph and uses BFS to find connected components\cite{gdbscan}.

\item[CUDA-DClust+]
CUDA-DClust+ \cite{9680379} uses the idea of incrementally growing clusters in parallel using chains from CUDA-DClust \cite{cuda-dclust} but reduces memory footprint and index structure build time. As CUDA-DClust+ is strictly better than CUDA-DClust, we only evaluate the former.
\end{description}

For all cases, we used the authors' original source code, with the only modifications being those necessary to get the code to run on our GPU system and to handle our inputs (with the exception of FDBSCAN, which uses an early traversal termination optimization to improve execution time for single runs. In this work, we focus on typical DBSCAN use cases where the user is expected to run DBSCAN multiple times with different parameter values. We go into more detail in Section~\ref{sec:early_term}).

We do not compare against Densebox approaches such as FDBSCAN-Densebox \cite{DBLP:journals/corr/abs-2103-05162}, HDBSCAN-Densebox algorithms \cite{bps-hdbscan,hdbscan}, as they are specialized to improve performance in datasets with very high density regions. In the absence of such regions, performance remains the same or is worse. We also do not compare our performance with Mr.Scan \cite{6877517} and BPS-HDBSCAN \cite{bps-hdbscan} as they are designed to cluster very large datasets (billion-point scale) and the incurred overhead is not amortized for smaller (thousand/million-point scale) datasets. We also do not report results against cuML's DBSCAN implementation as we were more than 3 orders of magnitude faster in all cases.  

We vary the $\varepsilon$ parameter (defined in Section~\ref{sec:dbscan}) and dataset size such that we include a wide range of clusters: a few large clusters, and many small clusters.
We also look at a case where no clusters are formed in a dense dataset in Section~\ref{sec:ngsim}. We do not report results from varying {\em minPts} (defined in Section~\ref{sec:dbscan}) as it did not provide any new insights. 
We report execution times averaged over 10 runs.

Overall, we find that RT-DBSCAN is consistently faster in almost all cases. In particular, RT-DBSCAN is more than 2.5x faster on larger datasets. For smaller dataset sizes (Section~\ref{sec:eval-small-datasize}), the performance difference between RT-DBSCAN and FDBSCAN is not as pronounced due to the non-negligible BVH build time of RT-DBSCAN. We elaborate on the impact of BVH build time in Section~\ref{sec:build_time}. 

\subsubsection{RT-DBSCAN Performance on Small Dataset Sizes}\label{sec:eval-small-datasize}
This section evaluates the four DBSCAN implementations on a small dataset ({\em 16K} points).
We found that both G-DBSCAN and CUDA-DClust+ ran out of memory on our GPU for more than {\em 100K} points. For this reason, subsequent sections will only compare against FDBSCAN.

Overall, we found that RT-DBSCAN outperformed other approaches in most cases for {\em 16K} points. As we kept decreasing the number of points in the dataset, we found that RT-DBSCAN was between 1.5x and 2x slower than FDBSCAN when the dataset size was less than 500.

\begin{figure}[t]
    \centering
    \includegraphics[width=0.45\textwidth]{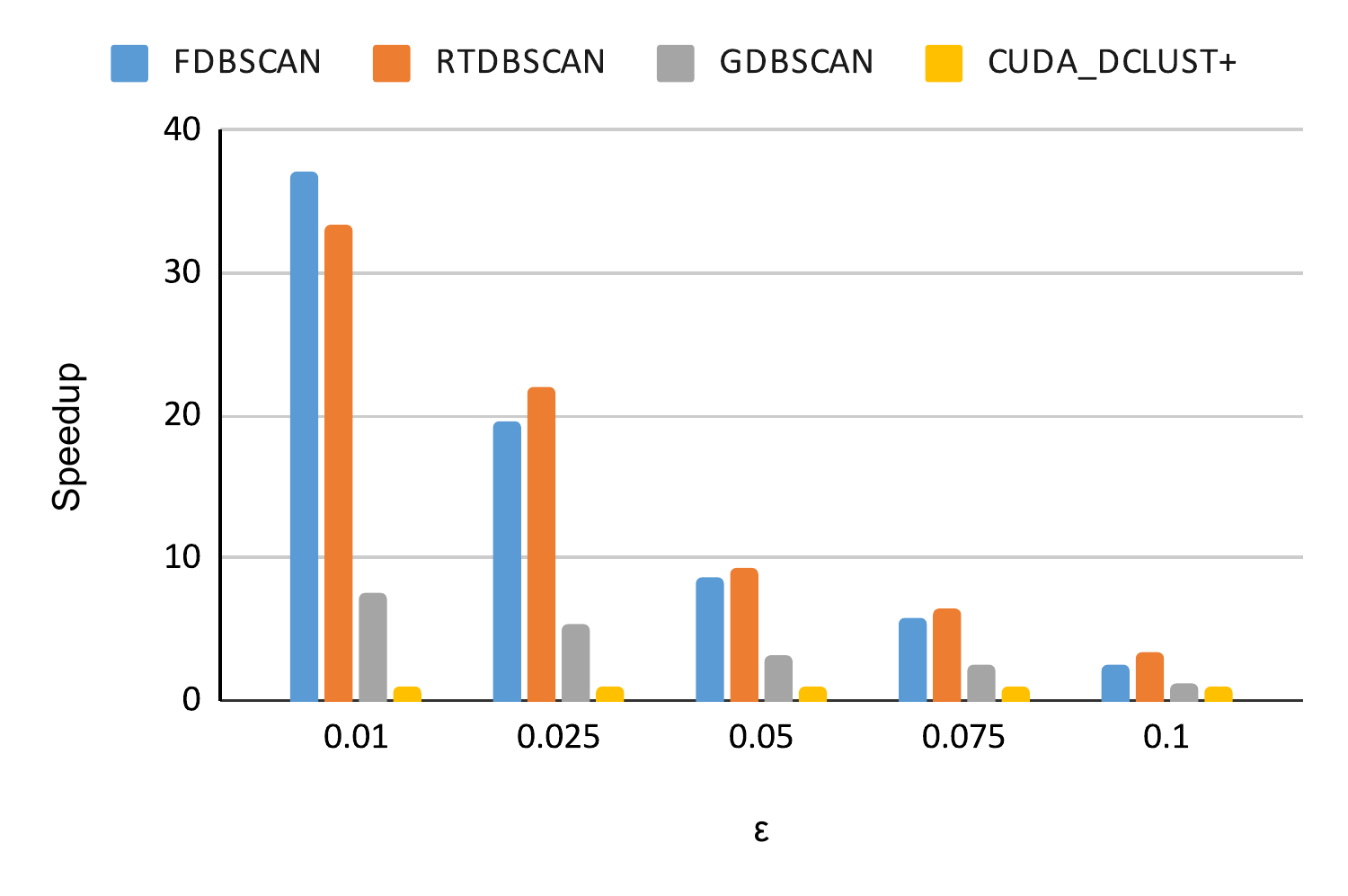}
    \vspace{-0.5em}
    \caption{Speedup over CUDA-DClust+ on varying search radius ($\varepsilon$) for {\em 16K} 3DRoad points}
    \label{fig:small-dataset-3droad}
    \vspace{-1.3em}
\end{figure}

We used {\em 16K} points from the 3DRoad dataset and set {\em minPts} as 100. In Fig~\ref{fig:small-dataset-3droad}, we compare the speedup of different approaches over CUDA-DClust+. It is evident that though RT-DBSCAN is faster in most cases, speedup is minimal compared to FDBSCAN, as the overhead of setting up the ray tracing framework was not amortized by the computations. We found that the poor performance of G-DBSCAN and CUDA-DClust+\footnote{We found that CUDA-DClust+ ran into memory issues on our 6GB GPU and also showed variability in clustering results between runs} was due to the time taken to traverse the adjacency list and the time needed to build and traverse the index structure, respectively.

\subsubsection{Impact of $\varepsilon$}\label{sec:eps}
We now turn to larger datasets, on which only FDBSCAN and RT-DBSCAN can run. We investigate clustering performance for different $\varepsilon$ values. We vary  $\varepsilon$ while fixing {\em minPts} as 100 and dataset size as {\em 1M}. We chose the first {\em 1M} points in the datasets for clustering and averaged our results over 10 runs.

\begin{figure*}
     \centering
     \begin{subfigure}[t]{0.32\textwidth}
         \centering
         \includegraphics[width=\textwidth]{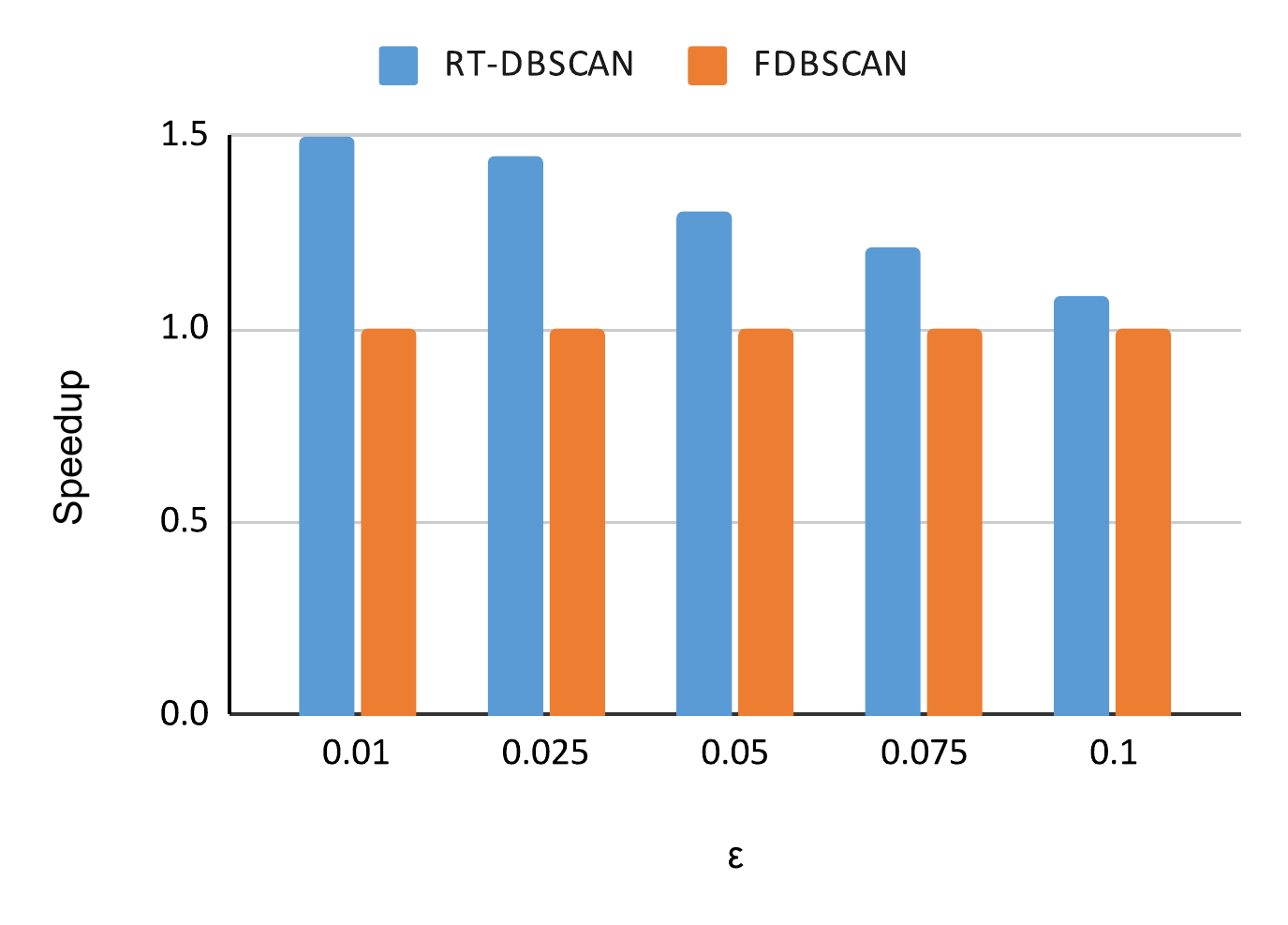}
         \caption{3DRoad}
         \label{fig:3droad_eps}
     \end{subfigure}
     \hfill
     \begin{subfigure}[t]{0.32\textwidth}
         \centering
         \includegraphics[width=\textwidth]{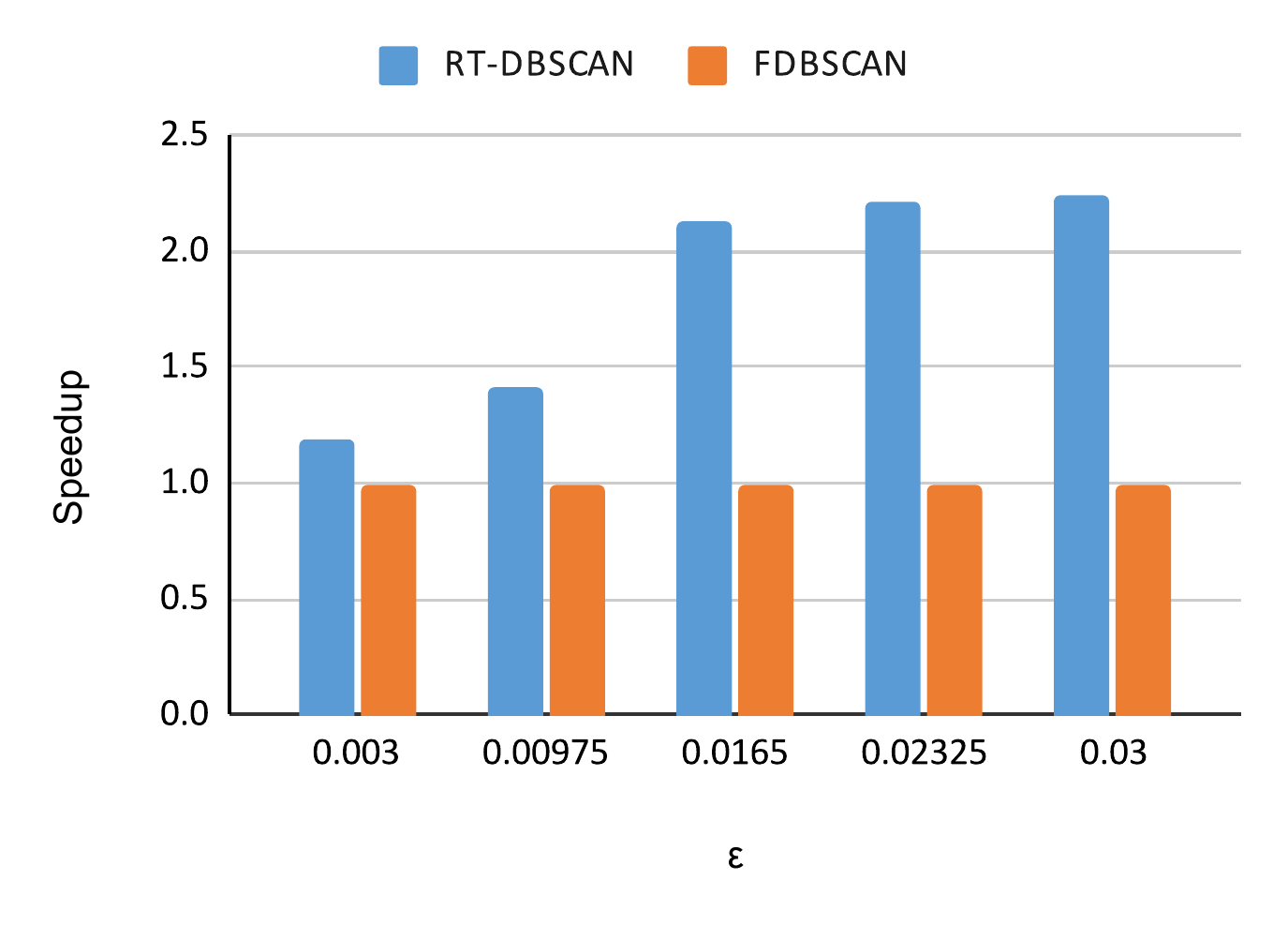}
         \caption{Porto}
         \label{fig:porto_eps}
     \end{subfigure}
     \hfill
     \begin{subfigure}[t]{0.32\textwidth}
         \centering
         \includegraphics[width=\textwidth]{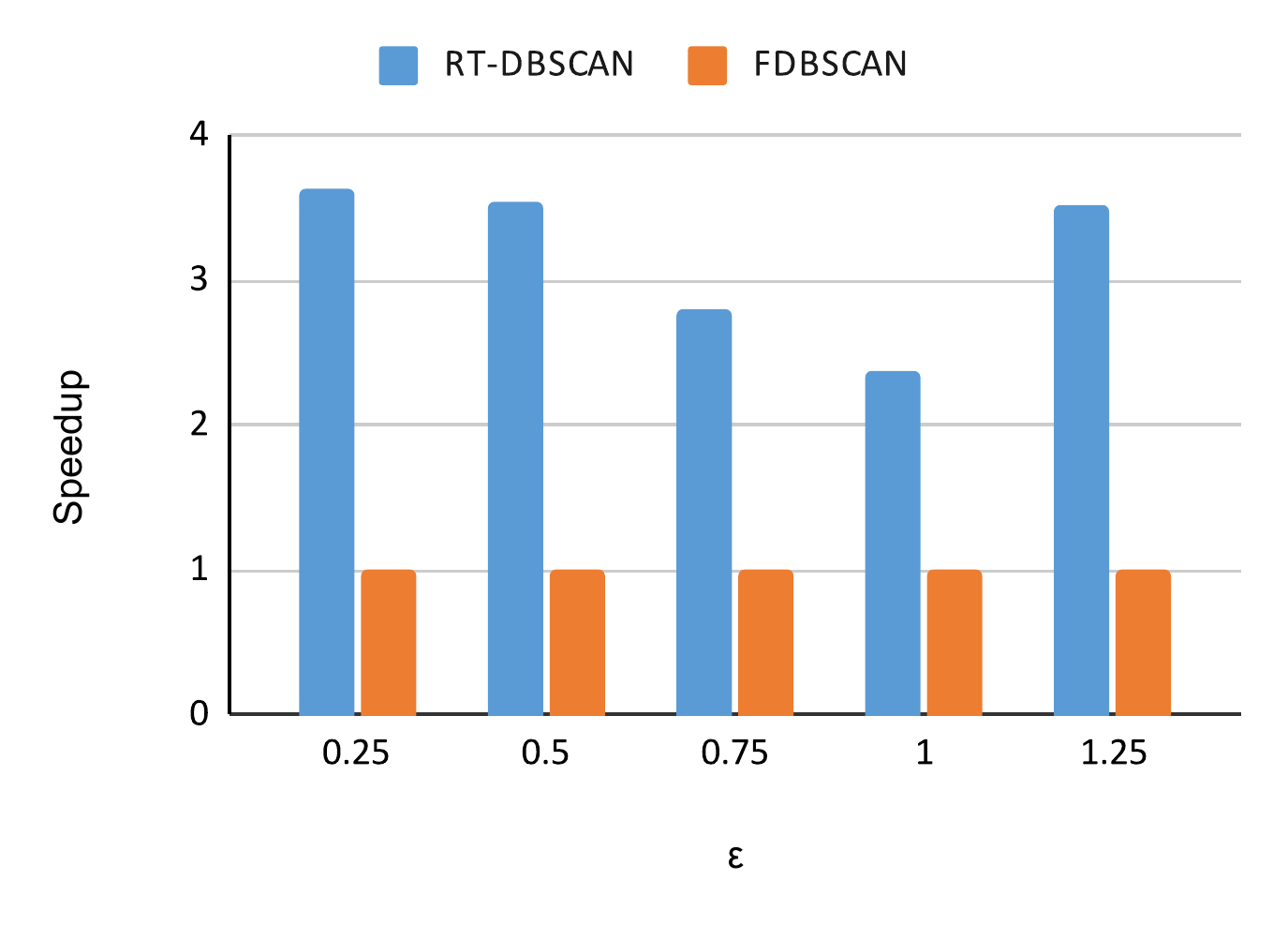}
         \caption{3DIono}
         \label{fig:3diono_eps}
     \end{subfigure}
        \caption{Speedup over FDBSCAN on varying search radius ($\varepsilon$)}
        \label{fig:varying_eps}
        \vspace{-1.4em}
\end{figure*}
We observe from Fig~\ref{fig:varying_eps} that RT-DBSCAN outperforms FDBSCAN in all cases. We attribute the speedup entirely to our ability to leverage hardware acceleration of BVH traversal, as FDBSCAN is also a BVH-based DBSCAN implementation, though it does not utilize RT cores. 

We see a maximum speedup of 1.5x on the 3DRoad dataset as shown in Fig~\ref{fig:3droad_eps}. As we will see in Section~\ref{sec:build_time}, DBSCAN execution time for small dataset sizes and small search radii is dominated by BVH build time.

In the cases of Porto and 3DIono, BVH build time of RT-DBSCAN was only 2.5x slower than FDBSCAN, allowing us to leverage the speedup in BVH traversal for fast clustering. For the Porto dataset in Fig~\ref{fig:porto_eps}, we see a maximum speedup of 2.3x and our speedup tended to increase with increasing $\varepsilon$ values. 

For the 3DIono dataset in Fig~\ref{fig:3diono_eps}, we achieve a maximum speedup of 3.6x. As the neighborhood search radius $\varepsilon$ becomes larger, the number of BVH traversals and intersection tests performed also increases, allowing us to realize the full potential of RT acceleration. 

\subsubsection{Impact of Dataset Size}\label{sec:dsize}
\begin{figure*}
     \centering
     \begin{subfigure}[t]{0.32\textwidth}
         \centering
         \includegraphics[width=\textwidth]{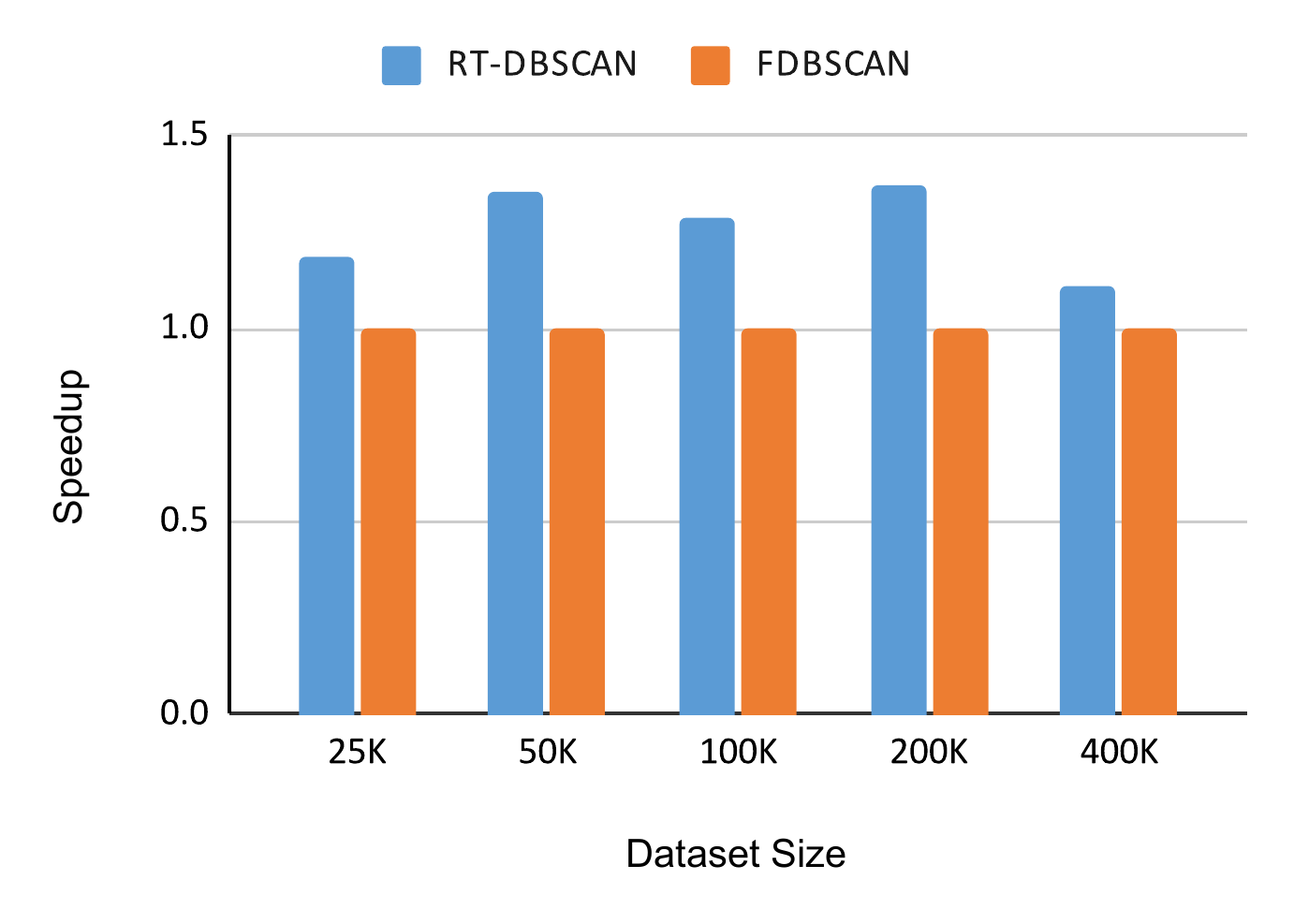}
         \caption{3DRoad}
         \label{fig:3droad_ds}
     \end{subfigure}    
     \hfill
     \begin{subfigure}[t]{0.32\textwidth}
         \centering
         \includegraphics[width=\textwidth]{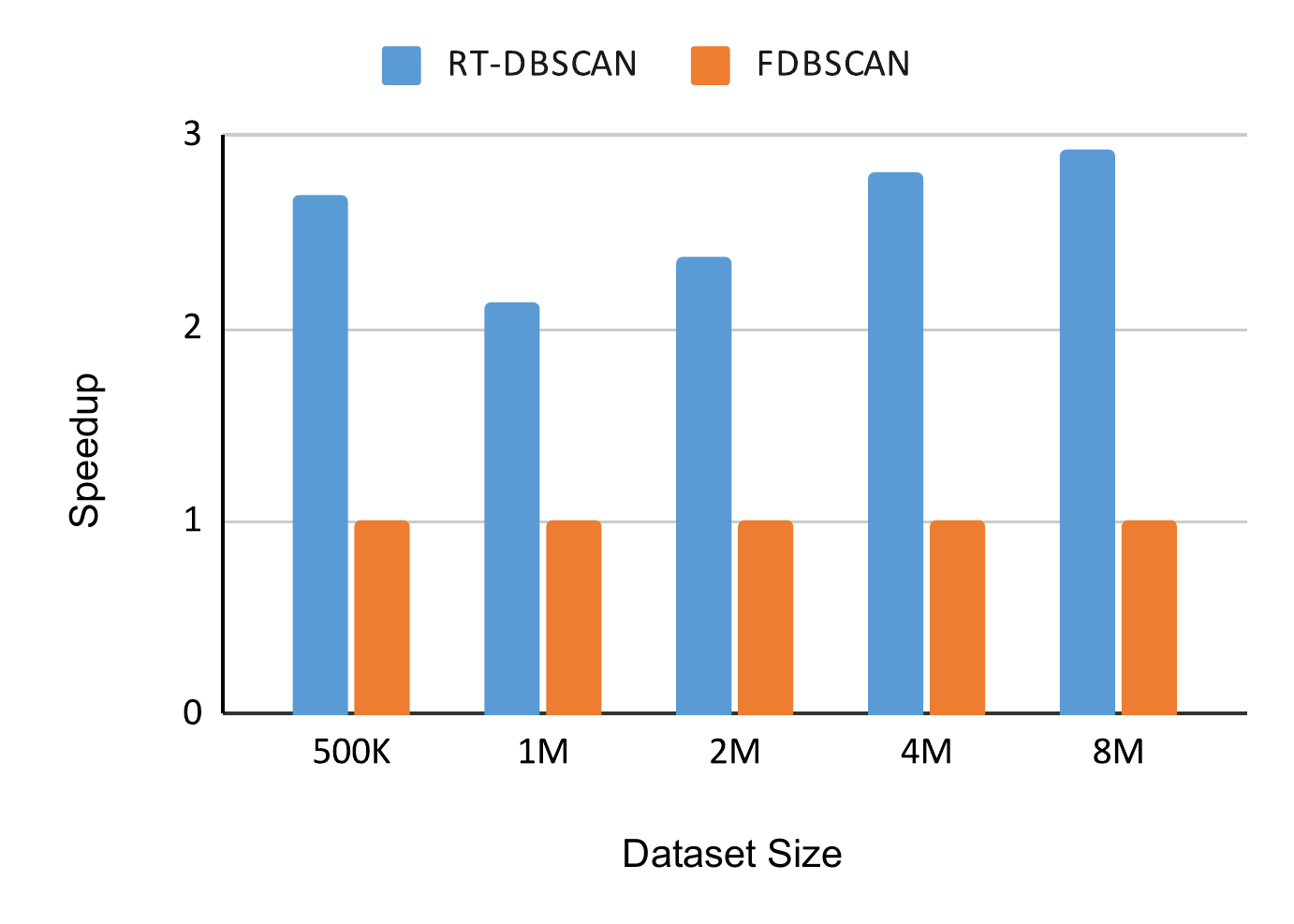}
         \caption{Porto}
         \label{fig:porto_ds}
     \end{subfigure}
          \hfill
     \begin{subfigure}[t]{0.32\textwidth}
         \centering
         \includegraphics[width=\textwidth]{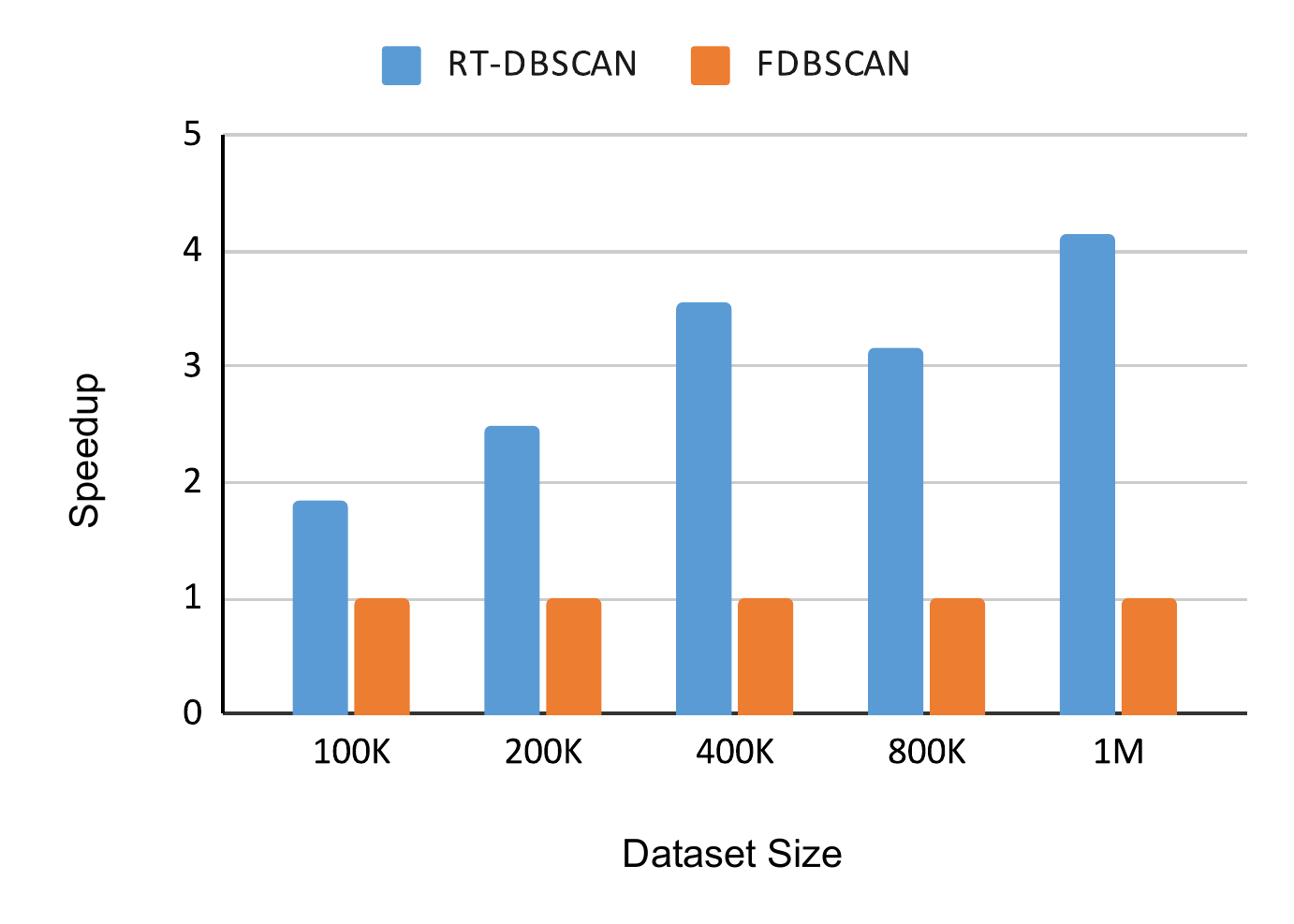}
         \caption{3DIono}
         \label{fig:3diono_ds}
     \end{subfigure}      
        \caption{Speedup over FDBSCAN on varying dataset size}
        \label{fig:varying_ds}
        \vspace{-1.3em}
\end{figure*}

Fig~\ref{fig:varying_ds} shows how performance of RT-DBSCAN and FDBSCAN varies with the size of the input dataset. We fix the ($\varepsilon$,{\em minPts}) values as (0.05,100), (0.5,10) and (0.5,1000) for the 3DRoad, 3DIono and Porto datasets respectively. For different dataset sizes ({\em n}), we choose the first {\em n} points for clustering.

We find that RT-DBSCAN outperforms FDBSCAN on all datasets and the performance disparity is especially evident for larger dataset sizes. For the 3DRoad dataset, we see from Fig~\ref{fig:3droad_ds} that our maximum speedup is 1.37x. As 3DRoad is  relatively small with a maximum of {\em 400K} points, it is not surprising that we face issues similar to those discussed in Section~\ref{sec:eps}, where BVH build time is not amortized by the time taken to complete the two stages of the DBSCAN algorithm.

\begin{figure}[t]
     \centering   
     \includegraphics[width=0.45\textwidth,scale = 0.5]{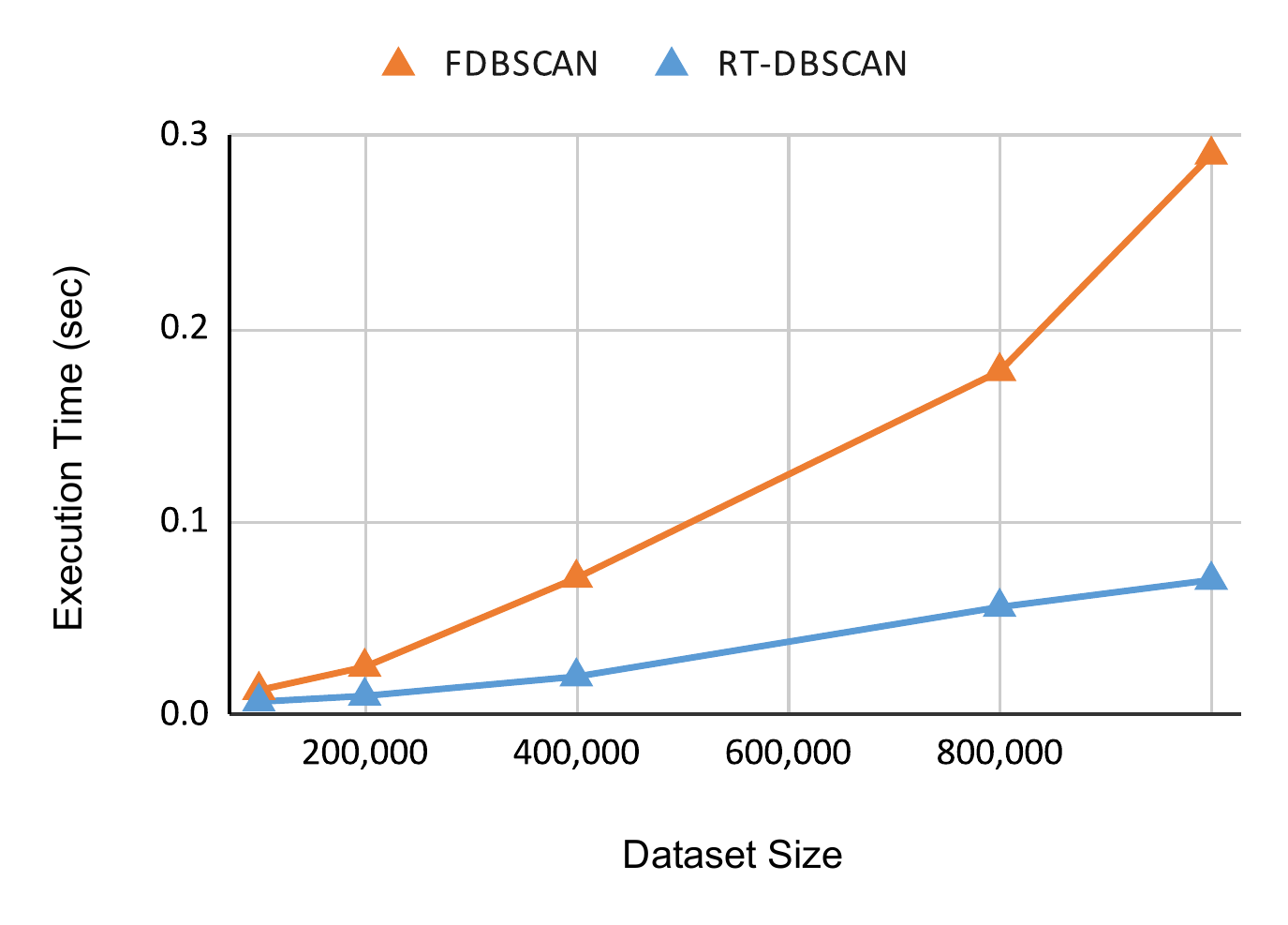}
     \vspace{-0.5em}
     \caption{Scalability of execution time for 3DIono dataset}
     \label{fig:scalable-exec-time}
\end{figure}


 For the Porto (Fig~\ref{fig:porto_ds}) and 3DIono (Fig~\ref{fig:3diono_ds}) datasets, we find that we achieve maximum speedups of 2.9x and 4.1x, respectively for the maximum dataset sizes. We report the raw execution time for Porto, the largest dataset we examined, in Table~\ref{table:porto-dataset-size}. 
 \begin{table}[h]
\centering
    \begin{tabular}{lcccl} 
    \toprule
       \textbf{Dataset size } & \textbf{FDBSCAN(s)} & \textbf{RT-DBSCAN(s)}\\
       \midrule
       500K & 539.85 & 200.82\\

       1M & 2868.1 & 1347.2\\

       2M & 14859.02 & 6264.6\\

       4M & 65935.14 & 23486.15\\

       8M & 282047.12 & 96333.7\\
       \bottomrule
   \end{tabular}
   \caption{Execution time (in seconds) for Porto dataset on varying dataset size}
    \label{table:porto-dataset-size}
     \vspace{-0.8em}
\end{table}
We examine the growth rate of the execution times of RT-DBSCAN and FDBSCAN on the 3DIono dataset in Fig~\ref{fig:scalable-exec-time}. We find that the growth rate of RT-DBSCAN's execution time is significantly slower than that of FDBSCAN as we are able to leverage hardware acceleration, showing that our approach is scalable. In general, increasing the dataset size widens the performance gap between RT-DBSCAN and FDBSCAN, as the RT hardware is designed to handle a large number of rays.

\subsection{RT-DBSCAN Performance on a Dense Dataset}\label{sec:ngsim}
Finally, we evaluated RT-DBSCAN on NGSIM, a very dense dataset where the number of clusters formed is 0, using the same criteria as in Sections~\ref{sec:eps} and \ref{sec:dsize}.

\begin{figure*}
     \centering
     \begin{subfigure}[t]{0.45\textwidth}
         \centering
         \includegraphics[scale=0.5]{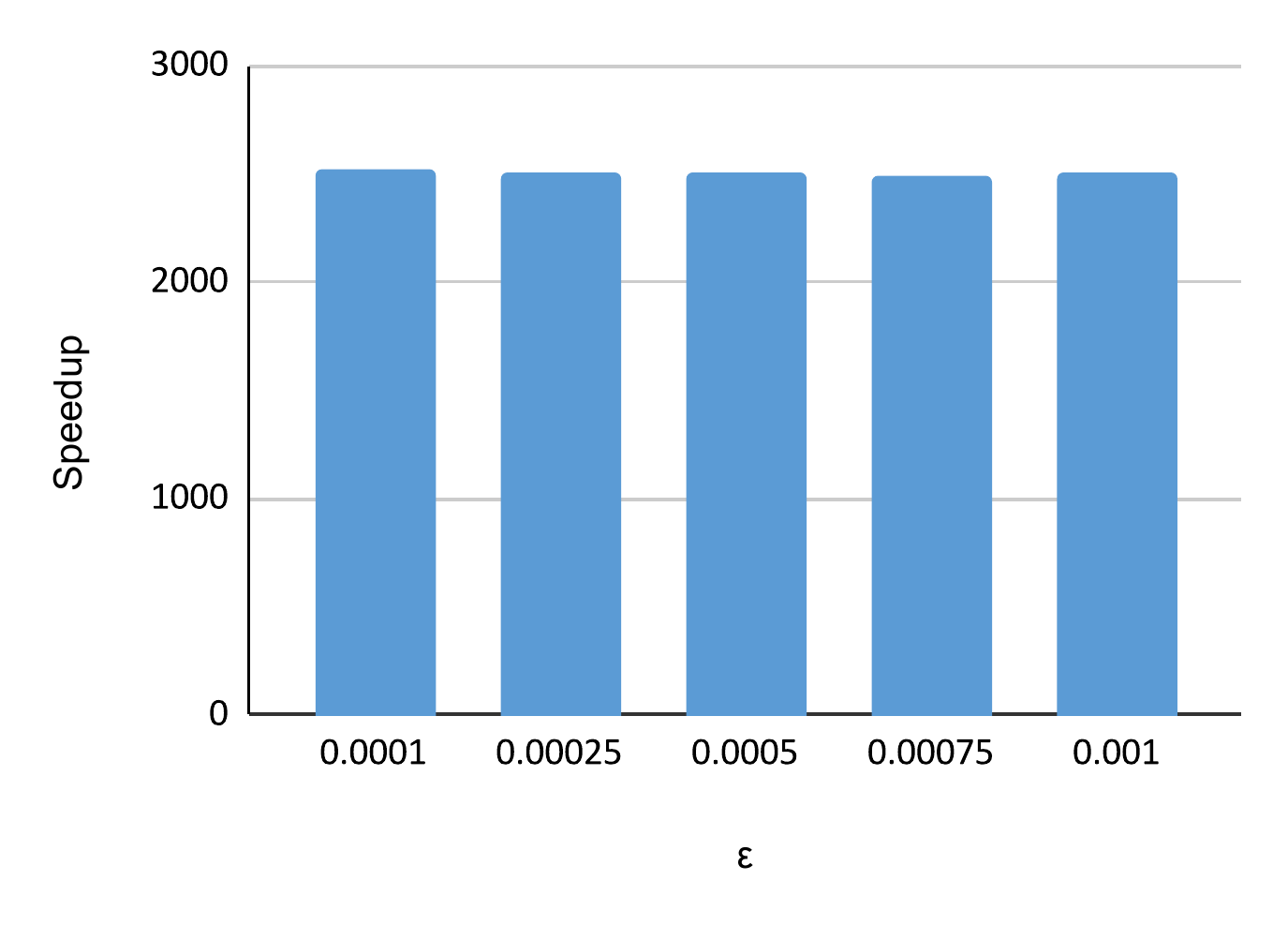}
         \vspace{-0.5em}
         \caption{Speedup on varying search radius ($\varepsilon$)}
         \label{fig:ngsim_eps}
     \end{subfigure}    
     \hfill
     \begin{subfigure}[t]{0.45\textwidth}
         \centering
         \includegraphics[scale=0.5]{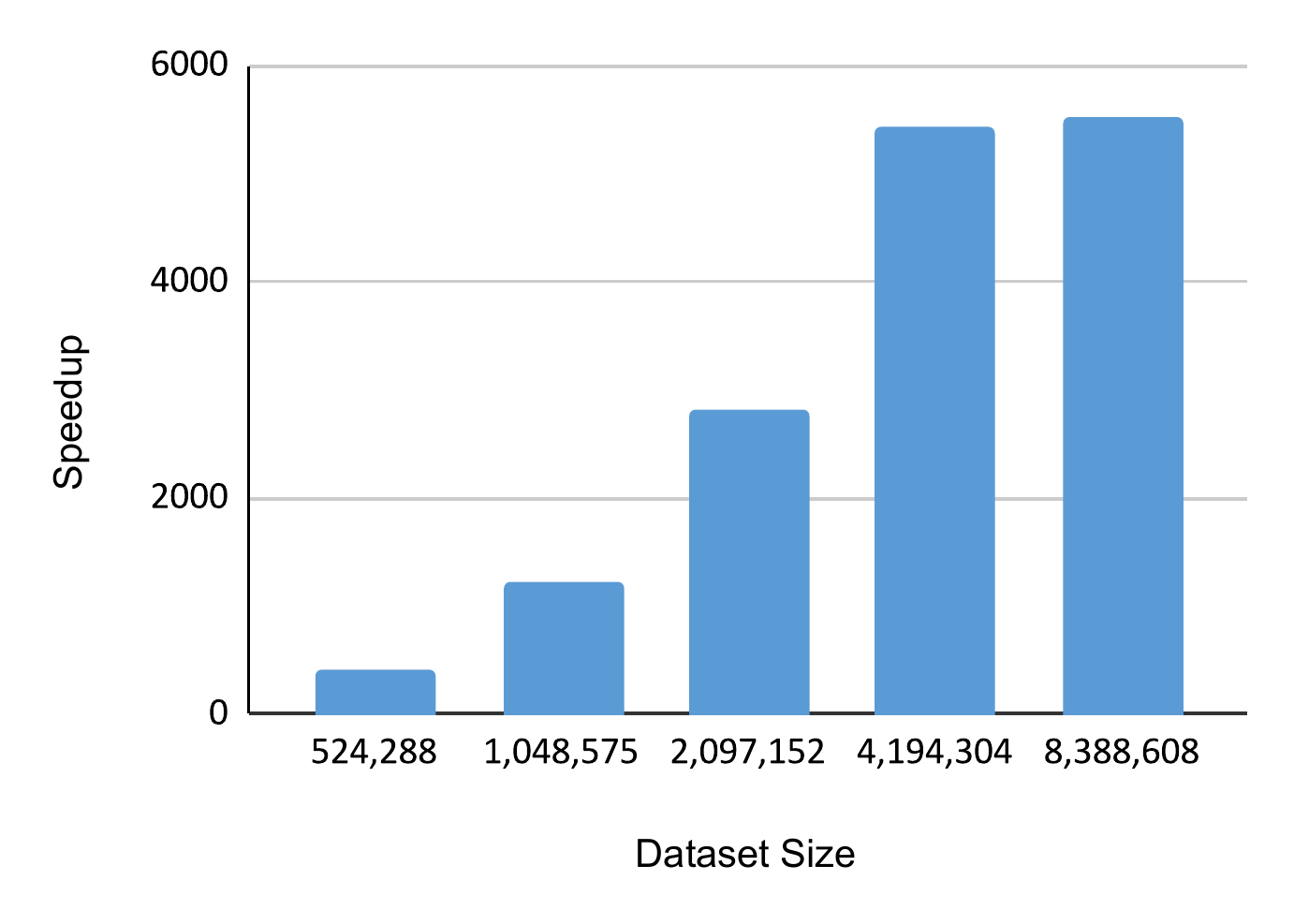}
         \vspace{-0.5em}
         \caption{Speedup on varying dataset size}
         \label{fig:ngsim_ds}
     \end{subfigure}
     \caption{Speedup over FDBSCAN on varying $\varepsilon$ and dataset size for NGSIM}
     \label{fig:ngsim}
     \vspace{-0.5em}
\end{figure*}

When we varied the dataset size, we found that RT-DBSCAN outperformed FDBSCAN by large margins, with a maximum of 5500x, as shown in Fig ~\ref{fig:ngsim_ds}. Table~\ref{table:ngsim-dataset-size} shows the raw execution times. 

\begin{table}[h]
\centering
    \begin{tabular}{lcccl} 
    \toprule
    \textbf{Search radius ($\varepsilon$) } & \textbf{FDBSCAN(s)} & \textbf{RT-DBSCAN(s)}  \\         
       \midrule
        0.0001 & 64.72 & 0.0257\\

        0.00025 & 64.77 & 0.0259\\

        0.0005 & 64.74 & 0.0259\\

        0.00075 & 64.71 & 0.026\\

       0.001 & 64.74 & 0.0259\\
       \bottomrule
   \end{tabular}
   \caption{Execution time (in seconds) for NGSIM dataset on varying search radius ($\varepsilon$)}
    \label{table:ngsim-eps}
     \vspace{-1em}
\end{table}

On varying $\varepsilon$ with {\em minPts} as 100 and dataset size as {\em 1M}, we found that RT-DBSCAN was nearly 2500x faster than FDBSCAN, as shown in Fig ~\ref{fig:ngsim_eps}. Table~\ref{table:ngsim-eps} shows raw execution times for different $\varepsilon$ values. The execution times of both FDBSCAN and RT-DBSCAN did not significantly change for different $\varepsilon$ as the dataset was still dense for these different radii. 
\begin{table}[h]
\centering
    \begin{tabular}{lcccl} 
    \toprule
       \textbf{Dataset size } & \textbf{FDBSCAN(s)} & \textbf{RT-DBSCAN(s)} \\
        \midrule
       500K & 12.7 & 0.03\\

       1M & 72.8 & 0.06\\

       2M & 364.6  & 0.13\\

       4M & 1631.4 & 0.3 \\

       8M & 6964.1 & 1.26\\
       \bottomrule
   \end{tabular}
   \caption{Execution time (in seconds) for NGSIM dataset on varying dataset size}
    \label{table:ngsim-dataset-size}
     \vspace{-1em}
\end{table}

On analyzing the output, we found that the RT hardware made relatively few calls to the intersection program.
As the specifics of BVH construction and traversal in RT hardware are unclear, we speculate that the hardware was able to construct the BVH such that we were able to prune large parts of the search space and minimize the number of intersection tests required.
As we will discuss in Section~\ref{sec:extend-optix-api}, having access to the workings of the hardware internals would help explain our results a lot better.

\subsection{Runtime Analysis of RT-DBSCAN}\label{sec:build_time}
In Section~\ref{sec:input_trans}, we discussed how data points are converted to spheres so that RT cores can build and traverse the BVH in hardware. 
Though this helps attain our objective of converting the nearest neighbor problem to a ray tracing query, it comes at a cost. Building a BVH from spheres for a ray tracing application is much more complex and time-consuming than building a spatial tree for data points.
The Optix builder performs memory compaction, invokes bounding box routines and other ray-tracing-specific operations that add to the BVH build time. 

The general trend we observed was that BVH build time dominated the total execution time for smaller datasets and cases where $\varepsilon$ was small, as fewer BVH traversals and intersection tests needed to be performed. For example, we consider the first 1 million points from 3DIono dataset with $\varepsilon$ = 0.25, and {\em minPts} = 100, similar to Section~\ref{sec:eps}. We found that RT-DBSCAN was 3.6x faster than FDBSCAN overall. Breaking down the execution time, we found that the time taken by RT-DBSCAN to perform clustering operations {\em after} BVH build was 6.4 ms for the Core point identification phase and 6.6 ms for the cluster formation phase. In total, RT-DBSCAN only spent 48\% of total execution time) on actual clustering operations. On the other hand, FDBSCAN spent 0.118 seconds (94\% of total execution time) on clustering operations. This shows us that, on average, RT-DBSCAN is more than 9x faster than FDBSCAN in performing the actual clustering operations! 

\section{Discussion}
\subsection{Hardware Limitations}
A major disadvantage of using RT cores to accelerate distance computations is that the dimensionality of the dataset can be at most three. Indeed, the RT cores themselves expect the dataset to be {\em exactly} three dimensions. Despite this limitation, we note that there are many important real-world 2D and 3D datasets such as Geospatial data, point clouds, and object geometries, and important distance algorithms such as DBSCAN, computing normals, and filtering point cloud noise that use distance searches over these datasets. Indeed, we note that most of the prior DBSCAN works evaluate their approaches on 2D geospatial datasets~\cite{cuda-dclust,DBLP:journals/corr/abs-2103-05162,dbscan-compare}. It is possible to reduce the number of dimensions in the dataset using dimensionality reduction techniques such Principal Component Analysis, though this introduces approximation.


\subsection{Impact of early traversal termination} \label{sec:early_term}
FDBSCAN uses an optimization where it stops BVH traversal when {\em minPts} neighbors are found in the {\tt FindNeighbors} function\cite{DBLP:journals/corr/abs-2103-05162}. However, the Optix API, based on which OWL is built, does not allow BVH traversal termination unless the {\tt Intersection} kernel makes an additional call to the {\tt AnyHit} kernel. As this can incur significant overhead, RT-DBSCAN does not attempt to perform early traversal termination. 

Though the early exit optimization works very well for cases where the user is only expected to run DBSCAN {\em once}, it is, in practice, more useful to record the number of neighbors of every point by allowing the BVH traversal to run its course. By saving the number of neighbors of each point, we do not have to re-run core point identification phase (Stage-1 of Algorithm~\ref{alg:parallel_dbscan}) for any subsequent DBSCAN runs where the user changes the {\em minPts} parameter. 

\begin{figure*}
     \centering
     \begin{subfigure}[b]{0.32\textwidth}
         \centering
         \includegraphics[width=\textwidth]{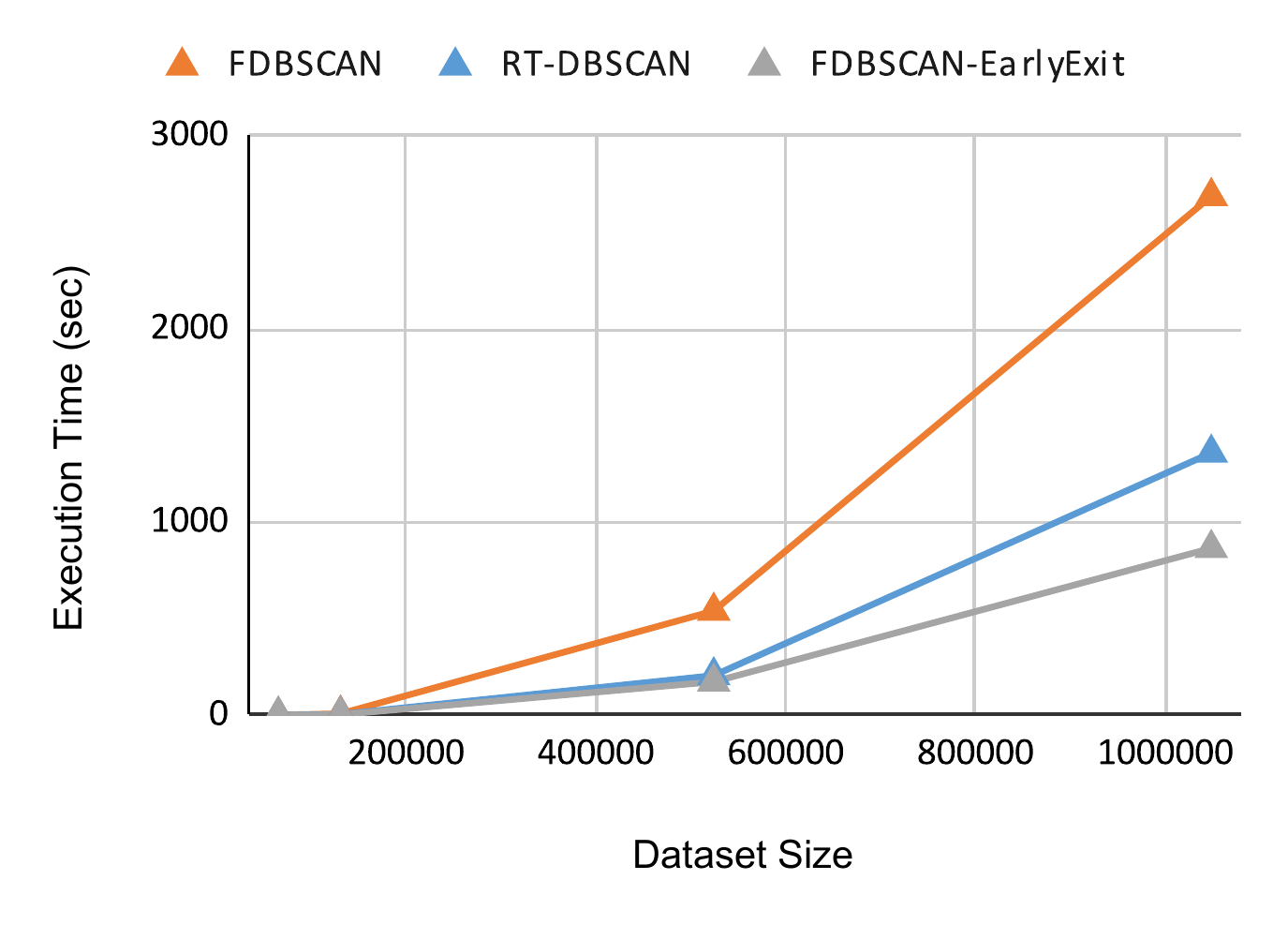}
         \caption{3DRoad}
         \label{fig:porto_early-exit}
     \end{subfigure}    
     \hfill
     \begin{subfigure}[b]{0.32\textwidth}
         \centering
         \includegraphics[width=\textwidth]{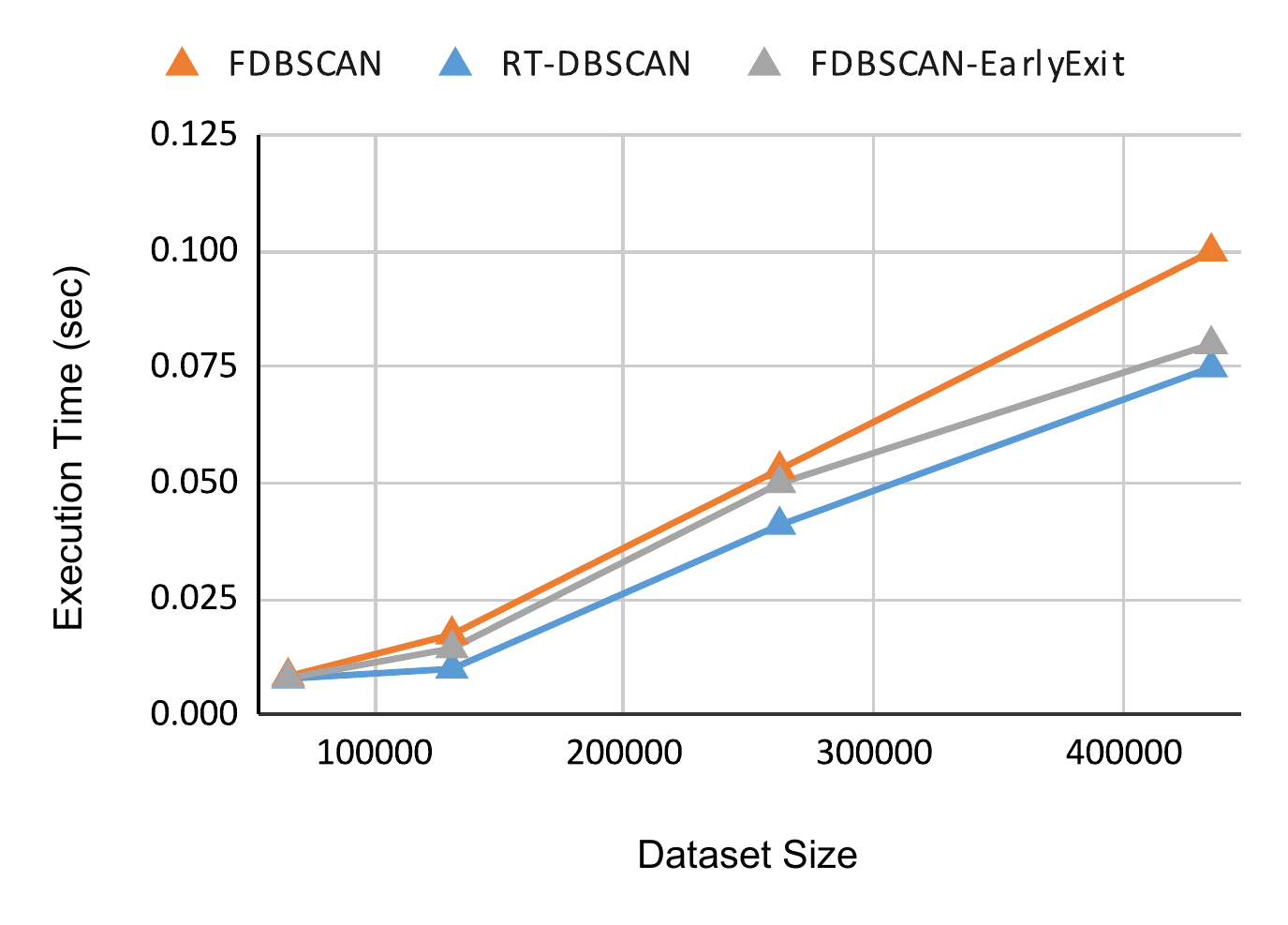}
         \caption{Porto}
         \label{fig:3droad_early-exit}
     \end{subfigure}
          \hfill
     \begin{subfigure}[b]{0.32\textwidth}
         \centering
         \includegraphics[width=\textwidth]{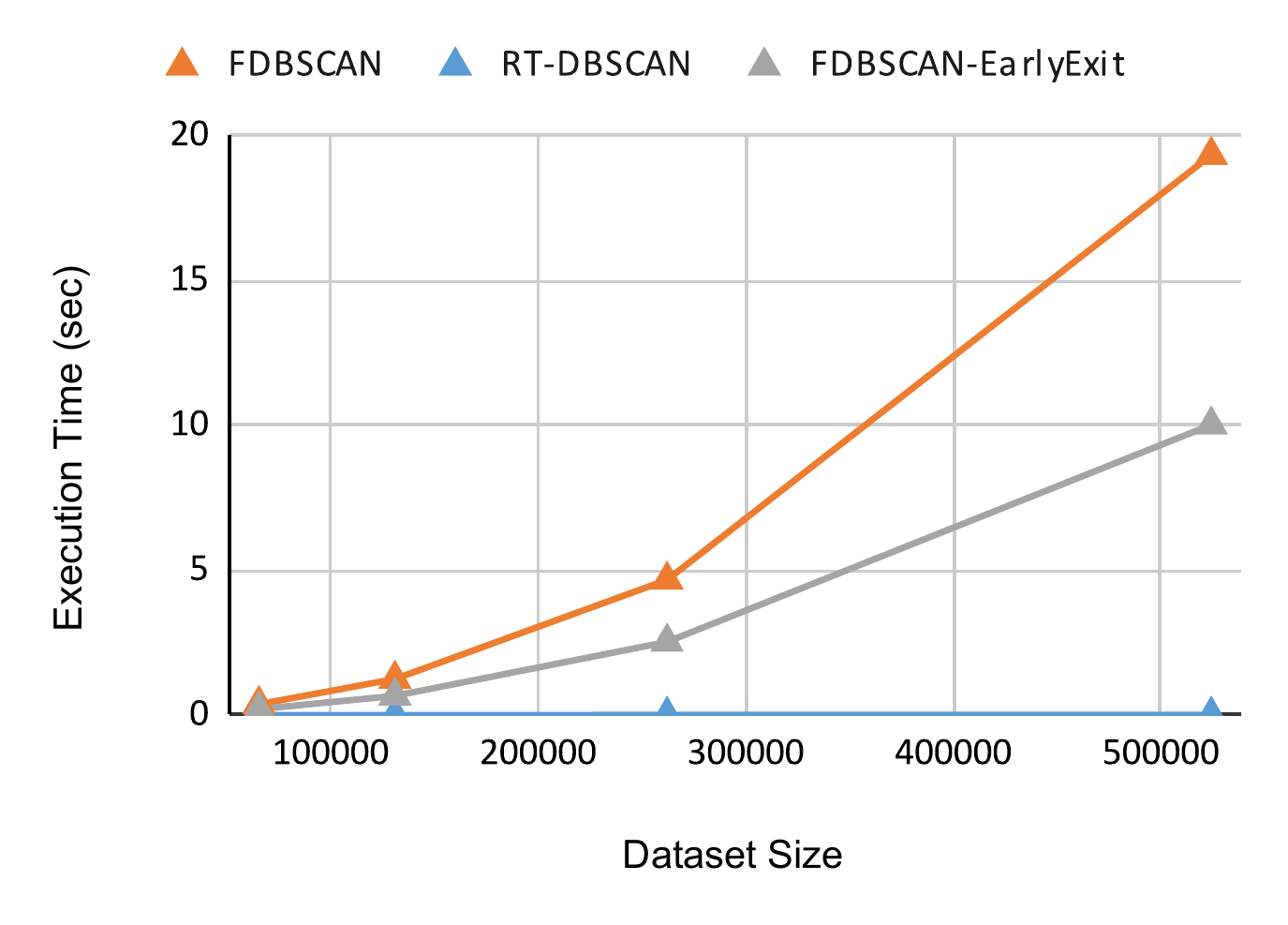}
         \caption{NGSIM}
         \label{fig:ngsim_early-exit}
     \end{subfigure}      
        \caption{Impact of early traversal termination on execution time}
        \label{fig:early-exit}
        \vspace{-1.5em}
\end{figure*}
In Figs~\ref{fig:early-exit}, we compare RT-DBSCAN to FDBSCAN {\em with} early termination, as well as FDBSCAN {\em without} while varying dataset size for fixed ({\em minPts}, $\varepsilon$) values. As expected, using early termination guarantees better performance as it often performs orders of magnitude fewer distance computations. This is especially true when {\em minPts} is very small and BVH traversal can stop very early. From our experiments, it is evident from Fig~\ref{fig:porto_early-exit} that for the Porto dataset, using early exit improves performance of FDBSCAN-EarlyExit by 3x compared to FDBSCAN {\em without} and by 1.5x compared to RT-DBSCAN for larger dataset sizes.

However, in other cases, even the early-exit optimization is not enough to overcome RT-DBSCAN's superior performance. For example,  note that RT-DBSCAN outperforms FDBSCAN-EarlyExit on the 3DRoad dataset (Fig~\ref{fig:3droad_early-exit}) and vastly outperforms it on the NGSIM dataset (Fig~\ref{fig:ngsim_early-exit}). We explain our findings by noting that the cost of additional intersection tests is hidden by the acceleration from RT cores. We especially note that even though the early exit optimization is able to leverage the density of the NGSIM dataset to improve performance substantially, RT-DBSCAN's ability to massively prune the search space is even more useful.

\subsection{Extensions to the Ray Tracing API}\label{sec:extend-optix-api}
We believe that it would be advantageous to have a certain degree of control over the hardware BVH traversal. For applications such as Barnes-Hut \cite{barnes86a}, it is necessary to access and update the intermediate nodes of the BVH to estimate the effect of the enclosed volumes and this is not possible with the current setup of the Optix API. We leave the Barnes Hut implementation as future work.

In this work, we leveraged the acceleration from the hardware BVH build and traversal. However, from Section~\ref{sec:rtx}, we know that the RT cores can also accelerate ray-triangle intersection tests. Indeed, Wald~\etal find that using hardware-accelerated intersection tests {\em in addition} to hardware-accelerated BVH traversal can produce substantial performance gains\cite{wald19}.

In our nearest neighbors algorithm (Algorithm~\ref{alg:rt-neigh}), we expand spheres around the points in the dataset and designate all points that fall within the sphere as neighbors of that point. We performed some experiments to see if we could approximate the spheres using triangles to leverage the hardware acceleration. As the ray-triangle intersection tests were done in hardware, we had to call the {\tt AnyHit} kernel to collect the intersected points. We found that using triangles resulted in 2x to 5x performance degradation due to the overhead cost associated with calls to the {\tt AnyHit} kernel. If there was an extension to the hardware such that the intersected points can be returned without using the costly {\tt AnyHit} kernel, there is massive potential for performance improvement from leveraging hardware-accelerated ray-triangle intersection testing.

\section{Related Work}
\paragraph{Using RT cores for non-Ray-Tracing Applications}
Wald~\etal first used RT cores to accelerate non-ray-tracing programs~\cite{wald19}. 
They formulated the problem of identifying a point's location in a tetrahedral mesh as a ray tracing problem by declaring the meshes as 3D objects in a scene and tracing rays originating at the query point. 
They show how leveraging both hardware BVH traversal and ray-triangle intersections resulted in upto 6.5x speedup over other CUDA implementations. 
Morrical~\etal used RT cores to successfully accelerate the unstructured mesh point location problem~\cite{Morrical2019EfficientSS}. Zellmann~\etal proposed a mapping of the fixed-radius nearest neighbor query to a ray tracing query for the Spring Embedders force-directed graph drawing algorithm\cite{forcegraph}.
Evangelou~\etal used the nearest neighbor mapping to solve the k-nearest neighbors problem\cite{Evangelou2021RadiusSearch}. Zhu proposed query re-ordering and partitioning algorithms to improve ray coherence and minimize the number of intersection tests performed\cite{rtnn}. We note that adding these optimizations to RT-DBSCAN would further improve performance.

\paragraph{DBSCAN}
Ester~\etal introduced the DBSCAN algorithm in 1996 to identify clusters of arbitrary shapes based on dense regions in the dataset\cite{Ester96adensity-based}. Although DBSCAN is an inherently sequential algorithm (Algorithm\ref{alg:dbscan}), researchers have exploited GPU parallelism to accelerate DBSCAN. Thapa~\etal exploited parallelism by having multiple CPU threads perform $\varepsilon$-neighbor distance computations in parallel\cite{5612134}. Andrade~\etal proposed G-DBSCAN, where they built a graph over the dataset and performed parallel Breadth First Searches to mark reachable points as belonging to the same cluster\cite{gdbscan}. 
However, the memory required to store and maintain the graph structure affects the scalability of G-DBSCAN. B\"{o}hm~\etal introduced CUDA-DClust, which used a spatial index structure to incrementally grow clusters in parallel\cite{cuda-dclust}. 
Poudel~\etal proposed CUDA-DClust+ which improved on CUDA-DClust by building the index structure on the GPU instead of CPU and reducing communication overhead\cite{9680379}. However, CUDA-DClust+ requires a significant amount of time for index construction and suffers when the size of GPU memory is small. Prokopenko~\etal proposed FDBSCAN and FDBSCAN-DenseBox that use Bounding Volume Hierarchy with UNION-FIND for clustering\cite{DBLP:journals/corr/abs-2103-05162}. Both FDBSCAN and FDBSCAN-DenseBox avoid memory issues as they do not store any neighbor information. FDBSCAN-DenseBox, similar to \cite{6877517,bps-hdbscan}, superimposes a Cartesian grid-based indexing to identify dense regions and reduce the number of distance computations in dense boxes.

\section{Conclusion}
In this work, we implemented RT-DBSCAN, where we accelerated the nearest neighbor searches in DBSCAN using Ray Tracing cores. We found that the hardware acceleration led to performance improvements by as much as 4.5x over current state-of-the-art GPU-based DBSCAN implementations. Future work entails removing the fixed-radius constraint for neighbor searches to accelerate a wider range of applications. It would also be interesting to see if RT cores can be used to accelerate more general tree traversal algorithms.

\balance
\bibliography{ref}
\bibliographystyle{IEEEtran}

\end{document}